\documentclass[sigconf]{acmart}

\usepackage{booktabs} 
\usepackage{graphicx}

\begin{document}
\title{PathRec: Visual Analysis of Travel Route Recommendations}

\author{Dawei Chen\raisebox{.5ex}{\small*$\dagger$}, Dongwoo Kim\raisebox{.5ex}{\small*}, Lexing Xie\raisebox{.5ex}{\small*$\dagger$},  
        Minjeong Shin\raisebox{.5ex}{\small*}, Aditya Krishna Menon\raisebox{.5ex}{\small$\dagger$*}, \\
        Cheng Soon Ong\raisebox{.5ex}{\small$\dagger$*}, Iman Avazpour\raisebox{.5ex}{\small$\ddagger$}, John Grundy\raisebox{.5ex}{\small$\ddagger$}}
 \affiliation{%
   \vspace{0.5\baselineskip}
   \institution{*The Australian National University, $\dagger$Data61/CSIRO, $\ddagger$Deakin University}
   \vspace{0.15\baselineskip}
 }
 \email{{u5708856, dongwoo.kim, lexing.xie, minjeong.shin, aditya.menon, chengsoon.ong}@anu.edu.au}
 \email{{iman.avazpour,j.grundy}@deakin.edu.au}

\renewcommand{\shortauthors}{}

\begin{abstract}
We present an interactive visualisation tool for recommending travel trajectories. 
This system is based on new machine learning formulations and algorithms for the sequence recommendation problem. 
The system starts from a map-based overview, taking an interactive query as starting point.  It then breaks down contributions from different geographical and user behavior features, and those from individual points-of-interest versus pairs of consecutive points on a route. The system also supports detailed quantitative interrogation by comparing a large number of features for multiple points. 
Effective trajectory visualisations can potentially benefit a large cohort of online map users and assist their decision-making.  
More broadly, the design of this system can inform visualisations of other structured prediction tasks, such as for sequences or trees. 

\end{abstract}


\maketitle
 
\section{Introduction}
Sequence recommendation has emerged as an important framework for modelling diverse problems such as travel route and music playlist recommendation~\cite{chen2017SR}. 
Unlike classical ranking algorithms where items are considered independently~\cite{koren2009matrix},
a sequence recommendation algorithm requires modelling a structure between items and suggests a set of items as a whole. 
For example, consider recommending a trajectory of \emph{points-of-interest} (POIs) in a city to a visitor. 
While a classical ranking algorithm can learn a user's preference for each individual location, it may ignore the distances between them and could suggest a longer trajectory than is optimal. 
Several sequence recommendation algorithms have been proposed to solve this problem and demonstrated superior performance compared to classical ranking algorithms~\cite{ijcai15,chen2017SR}. 
Nonetheless, recommendation algorithms for sequences and trajectories~\cite{chen2016learning,chen2017SR} have many components and can be difficult for a user to understand. This is part of the general challenge of introducing transparency and accountability for machine learning algorithms~\cite{fatml}. 

In this paper, we tackle the problem of sequence visualisation, specifically focussing on travel routes recommendation. 
A travel route is a sequence of POIs, and the sequence recommendation problem can be formulated as a structured prediction problem~\cite{chen2017SR}.
Based on a diverse set of features for individual and pairs of POIs, we train the prediction model with trajectory data extracted from geo-tagged photos taken in Melbourne~\cite{chen2016learning}. 
To visualise the suggested routes, we develop a novel tool that efficiently displays multiple suggested routes, which helps users understand the process behind the recommendations.
Specifically, our system decomposes a total score of each route into a set of features and their corresponding scores, and shows the total score as a stacked bar plot of the features.
The system also visualises the differences between POIs in a single route to show how POIs in that route can exhibit vast diversity. 

This visualisation helps tourists who want 
diverse experiences by choosing the best route among multiple recommendations. 
Generalising to a broader class of routes, such a visualisation could also help users of online mapping apps to make decisions on suggested travel routes, such as by trading off distance, traffic, and scenery.

 \begin{figure}
 \centering
   \includegraphics[width=\linewidth]{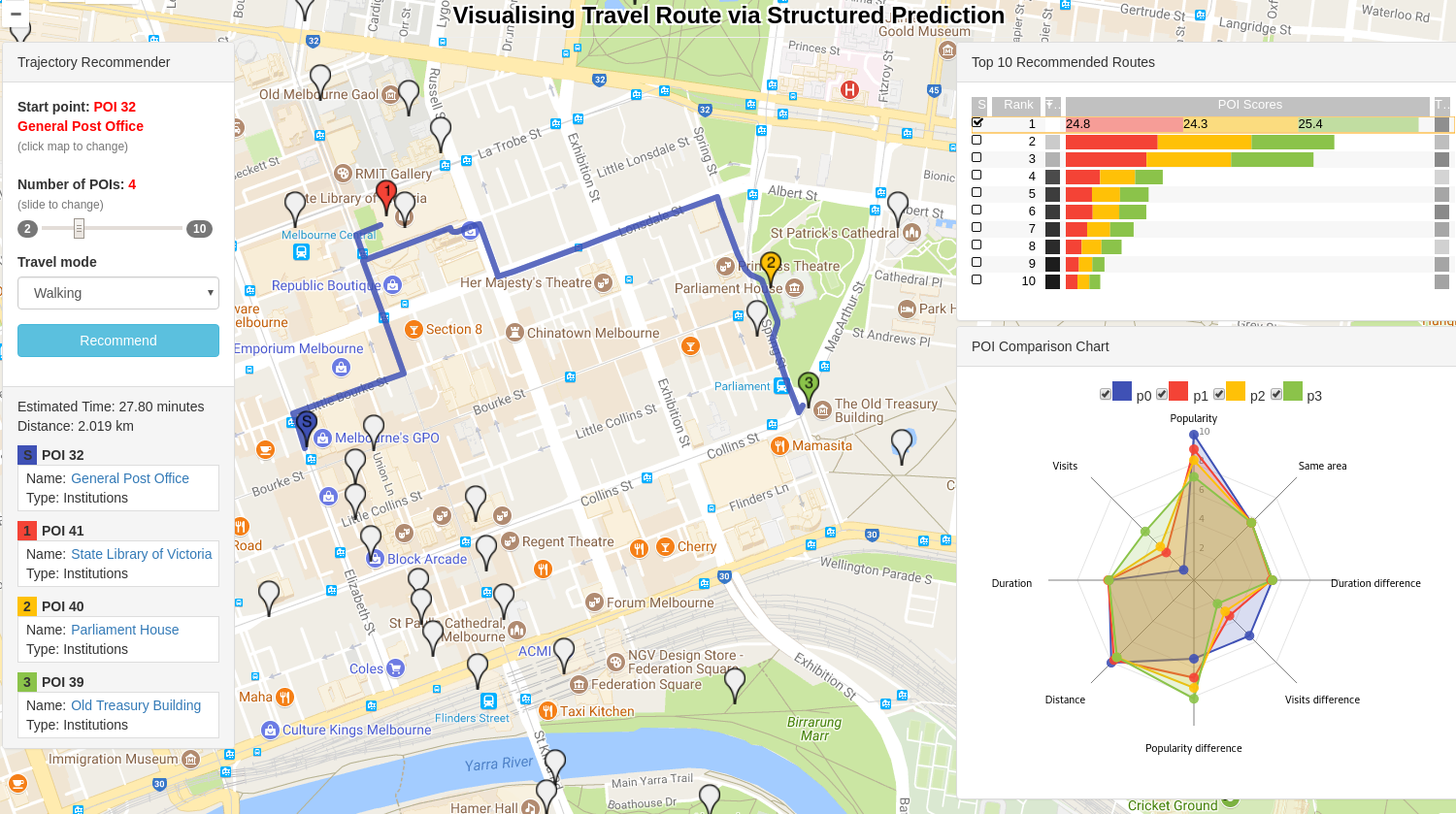} 
     \caption{Travel route visualisation system\protect\footnotemark. 
     Given a starting POI and the number of POIs to be visited, the system recommends multiple routes from travel history of tourists. Shown above: recommendation in central Melbourne.}
     \label{fig:overview}\vspace{-1em}
 \end{figure}

\footnotetext{\url{http://www.pathrec.ml}}
 
\section{Travel Route Recommendation}
The travel route recommendation problem involves a set of POIs in a city. 
Given a trajectory query $\mathbf{x} = (s, l)$, comprising a start POI $s$ and trip length $l$, the goal is to suggest one or more sequences of POIs that maximise some notion of utility.

Following~\cite{chen2017SR}, we first cast travel recommendation as a structured prediction problem, which allows us to leverage the well-studied literature of structured SVMs (SSVM)~\cite{joachims2009predicting}. 
From a visualisation perspective, an advantage of the SSVM is the explicit representation of feature scores in its final decision process. Specifically, we can disassemble the final score of a route into feature scores of each POI and the transition between two adjacent POIs. 
We use hand-crafted POI features such as the category, popularity, and average visit duration of previous tourists. We also crafted transition features such as the distance and neighbourhood of two POIs to maximise the interpretability of the outcome.

\begin{figure}[t!]
\includegraphics[width=\linewidth]{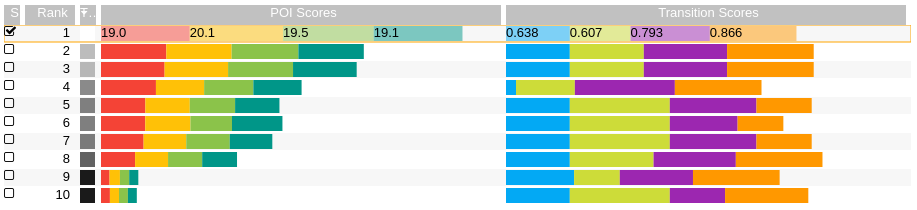}
\caption{Visualisation of POI and transition scores for top 10 recommended routes. Each bar from left to right represents a relative score of each POI or transition along the route.
The length of stacked bars represents the total score of the suggested route.}
\label{fig:stack} 
\end{figure}

\section{Visualisation}
Our goal is to design an interactive visualisation system on top of the structured prediction framework.
Figure~\ref{fig:overview} shows the overview of a live demo system, which consists of five major components: a map to display the suggested routes, an input box for user query (upper left), a stacked score of routes (upper right), a POI list box (lower left), and a radar chart to compare features of multiple POIs (lower right). 
The role and the construction of the four major components, besides the main map, are as follows:

\textbf{Query input}: A query consists of a starting POI and a trip length. 
Users can choose the starting POI by clicking icons on the map and can adjust the slide to set the trip length. 
In addition, three different travelling modes (e.g. bicycling, driving and walking) are supported, 
and we optimise the suggested routes for each mode.

\textbf{Route score visualisation}: 
The SSVM evaluates relevance scores of POIs and transitions in a candidate route to the given query and uses the sum of the relevance scores to determine the ranks of the routes.
To visualise the POI and transition scores, we adopt a stacked bar representation~~\cite{gratzl2013lineup}, designed to support the visualisation of multi-attribute ranking.
In Figure~\ref{fig:stack}, the system decompose the scores of top 10 recommended routes into POI and transition scores via the stacked bar representation, where the size of each bar is proportional to the relevance score of the corresponding POI and transition in the route.
Note that the POI and transition scores are scaled differently to support better visual discrimination\footnote{See Appendix for details at \url{http://arxiv.org/abs/1707.01627}}.
For a seamless match between a route on the map and the corresponding POI scores in the bar plot,
we use the same colour for both POI score and POI icon on the map.
We also allow users to select multiple rows to visualise the corresponding routes on the map.

\textbf{POI list}:
The POI list box provides the list of POI names and categories along the recommended route.
The list is sorted according to the suggested visiting order, and again, the same POI colour is used to match the corresponding POI on the map.
On top of the list, the system also provides an estimated travel time and total distance of the route. 
The POI list box is updated whenever a user selects a different route or the system makes a new recommendation. 
If more than one route is selected, the system displays the information of the most recent chosen route.

\begin{figure}[t!]
\includegraphics[width=0.6\linewidth]{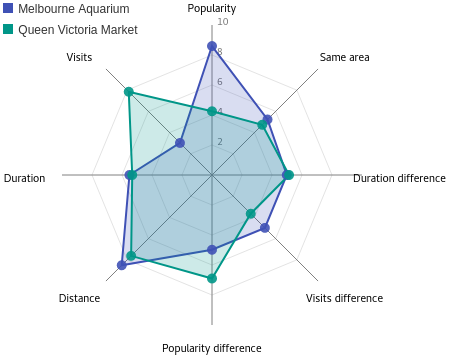} 
    \caption{POI feature comparison between \textit{Melbourne Aquarium} and \textit{Queen Victoria Market}: the former scores higher on \textit{Popularity} and \textit{Visits difference} features whereas the latter scores higher on \textit{Visits} and \textit{Popularity difference} features.}
\label{fig:radar} \vspace{-1em}
\end{figure}

\textbf{POI feature visualisation}: We further provide a radar chart to analyse the variation between POIs in a single route. 
For example, in Figure~\ref{fig:radar}, we compare two POIs (\textit{Melbourne Aquarium} and \textit{Queen Victoria Market}) in terms of POI features and their importance in the suggested route. 
The radar chart shows the corresponding POI feature scores when a user selects a route.
In particular, the user can check/uncheck any POI in the selected route, and the feature scores of all checked POIs will be shown in the chart.

\section{Conclusion}
In this demonstration, we detail an interactive route analyser which helps the interaction between users and a route recommendation system.
The system benefits from the explicit feature construction of the structured prediction model, and visualises recommended routes in terms of information on both the routes and the POIs.

{
\setlength{\parindent}{0cm}
\paragraph{\bf Acknowledgements}
We thank the National Computational Infrastructure (NCI), supported by the Australian Government, for computational resources.
This work is supported in part by the Australian Research Council via Discovery Project program DP140102185.
}
\vspace{-1em}


\onecolumn
\appendix

\section{Alternative approaches to trajectory recommendation}
\label{sec:alternative}

A number of approaches have been proposed to solve the trajectory recommendation problem.
\citet{ijcai15} formulated an optimisation problem inspired by the travelling salesman problem,
and \citet{cikm16paper} proposed to learn a RankSVM~\cite{lranksvm} model to rank POIs with respect to a query,
in particular, 
the training objective is
\vspace{-0.5em}
\begin{equation}
\label{eq:ranksvm}
\min_{\mathbf{w}} \frac{1}{2} \mathbf{w}^\top \mathbf{w} + C \cdot \sum_{i=1}^n \sum_{(p,p') \in \mathcal{R}(\mathbf{x}^{(i)})}
          \max\left(0, 1 - \mathbf{w}^\top \left(\Phi(\mathbf{x}^{(i)}, p) - \Phi(\mathbf{x}^{(i)}, p') \right) \right)^2,
\end{equation}
where $\mathbf{w}$ denotes the model parameters, $\Phi$ is a query-POI feature mapping, $C > 0$ is a regularisation constant,
and $\mathcal{R}(\mathbf{x})$ is the set of POI pairs $(p, p')$ such that $p$ is ranked above $p'$,
e.g. POI $p$ is observed more often than POI $p'$, with respect to query $\mathbf{x}$.
Lastly, the top-ranked POIs with respect to the given query were taken to form a trajectory.

Instead of ranking POIs, a Markov chain was learned from routes in the travel history~\cite{cikm16paper},
and recommending a trajectory is achieved with either the classic Viterbi algorithm or 
an integer linear programming when repeated POI visits are discouraged.
Further, \citet{cikm16paper} proposed to combine the POI ranks learned by RankSVM~(\ref{eq:ranksvm}) and 
the transition preferences learned by a Markov chain using a heuristic which traded off between point affinity and transition preference.

\section{Score normalisation for visualisation}
\label{sec:scorenorm}

We visualise multiple suggested trip by ranking them according to their scores from SSVM, in addition,
we leverage the linear form of SSVM that score of a trajectory is the sum of point and transition scores in the trip.
As shown in Figure 2, scores of both POIs and transitions in a trip are visualised in a stacked bar plot,
which helps users of the system to better understand what contributes to a suggested trip.

To support better visual discrimination, we perform linear scaling on scores of suggested trajectories, scores of POIs and transitions.

\subsection{Scaling overall trajectory scores}
Specifically, we scale the trajectory scores for the top 10 suggested routes such that:
\begin{itemize}
\item The first trajectory scores 100,
\item and the last (i.e. the 10-th) trajectory scores 10.
\end{itemize}

\subsection{Displaying unary and pairwise scores}
We further scale the POI scores using the same scaling parameters as that of trajectory scores, however, 
to visualise the transition scores in the stacked bar plot,
we perform another linear scaling as the scores of POIs and transitions are in quite different range,
in particular, the transition scores are scaled linearly into the range [0.1, 1].

\subsection{Displaying POI features}
Besides scores of POIs and transitions of a suggested trip, we also visualise the scores for individual 
POI features (e.g. popularity, visit duration, category) in a radar chart, as shown in Figure 3.
We perform another linear scaling such that scores of POI features are in the range [1, 10].

\section{Choices on visualisation paradigms}

We want to visualise a ranked list of suggested trips instead of a single choice, 
further, we would like to break down the score of a trip into the contributions of its unary and pairwise components.
Following the Visual Information-Seeking Mantra~\cite{shneiderman1996eyes},
we made the following design choices for this visualisation system:
{
\setlength{\parindent}{0cm}
\paragraph{\bf Overview first}
For a given query, as shown in Figure 1, 
the visualisation system shows the top-ranked trip on a map.
The other components of the system, as described in Section 3, 
can display more details of these recommendations after a user zooms in, as described below.

\paragraph{\bf Zoom and filter}
The system shows the scores of top-10 suggested trajectories and the scores of the corresponding POIs and transitions in a stacked bar plot,
as shown in Figure 2.
Further, a POI list box at the lower left side of the system interface 
displays the information of POIs including names and categories of the selected trip, as described in Section 3. 
Lastly, at the lower right side, contributions of individual features of all POIs in the selected trip are compared, 
and Figure 3 shows an example which compares individual feature scores between two POIs.

\paragraph{\bf Details-on-demand}
The system also support display and compare multiple suggested trajectories on map,
this is achieved by select the trips at the left side of the stacked bar plot (Figure 2).
All selected trips will be displayed on the map, which helps user choose the best trip among the multiple recommendations.
Moreover, comparison of individual features between multiple POIs in a trip is also supported,
in the radar chart, users can check one or more POIs in the most recent choose trip, 
the feature scores of all checked POIs will then be displayed and compared, using the same colour as the POI icon on the map.
}

We note that to visualise POI feature contributions, instead of radar chart, parallel coordinates~\cite{parallel_coord} is another viable option,
although both approaches have their own advantages and drawbacks~\cite{robbins2005creating}.

\end{document}